\documentclass[aps,twocolumn,preprintnumbers,%
superscriptaddress,floatfix]{revtex4}
\usepackage[dvips]{graphics}
\usepackage{amsmath,amsfonts,bm}
\usepackage{epsfig}
\begin{document}
\newcommand{\be}{\begin{eqnarray}}
\newcommand{\ee}{\end{eqnarray}}
\def\lsim{\mathrel{\rlap{\lower3pt\hbox{\hskip1pt$\sim$}}
     \raise1pt\hbox{$<$}}} 
\def\gsim{\mathrel{\rlap{\lower3pt\hbox{\hskip1pt$\sim$}}
     \raise1pt\hbox{$>$}}} 
\def\N{${\cal N}\,\,$}
\def\prl{Phys. Rev. Lett.}
\def\np{Nucl. Phys.}
\def\pr{Phys. Rev.}
\def\pl{Phys. Lett.}
\def\la{\langle}\def\ra{\rangle}
\def\del{\partial}
\def\calL{\cal L}\def\calK{\cal K}
\def\hatn{\hat{n}}\def\Amu{{\cal A}_\mu}\def\A{{\cal A}}
\newcommand\<{\langle}
\renewcommand\>{\rangle}
\renewcommand\d{\partial}
\newcommand\LambdaQCD{\Lambda_{\textrm{QCD}}}
\newcommand\Tr{\mathrm{Tr}\,}
\newcommand\+{\dagger}
\newcommand\g{g_5}
\def\bi{\bibitem}

\title{Hidden Local Symmetry and Dense Half-Skyrmion Matter}
\author{Mannque Rho}
 \affiliation{ Service de Physique
Th\'eorique,  CEA Saclay, 91191 Gif-sur-Yvette C\'edex, France\\
(E-mail: mannque.rho@cea.fr)}

\begin{abstract}
Transition from baryonic matter to color-flavor-locked quark matter is described in terms of skyrmion matter changing into half-skyrmion matter. The intermediate phase between the density $n_p$ at which a skyrmion turns into two half skyrmions and the chiral transition density $n_c^{\chi SR}$ at which hadronic matter changes over to quark matter  corresponds to a chiral symmetry restored phase characterized by  a vanishing quark condensate and a  {\em non-vanishing} pion decay constant. When hidden local fields are incorporated,  the vector manifestation of Harada-Yamawaki HLS theory implies that as density approaches $n_c^{\chi SR}$,  the gauge coupling $g$ goes to zero (in the chiral limit) and the symmetry ``swells" to $SU(N_f)^4$ as proposed by Georgi for the ``vector limit." This enhanced symmetry, not present in QCD,  can be interpreted as ``emergent" in medium due to collective excitations. The fractionization of  skyrmions into half-skyrmions resembles closely the magnetic N\'eel--to-valence bond solid (VBS) paramagnet transition where ``baby" half-skyrmions enter as relevant degrees of freedom in the intermediate phase. It is suggested that the half-skyrmion phase in dense matter corresponds to the ``hadronic freedom" regime that plays a singularly important role in inducing kaon condensation that leads to the collapse of massive compact stars into black holes..

\end{abstract}

\date{\today}

\newcommand\sect[1]{\emph{#1}---}

\maketitle

\sect{Introduction:}  Hadronic matter at high density is presently poorly understood and the issue of the equation of state (EOS) in the density regime appropriate for the interior of compact stars remains a wide open problem. Unlike at high temperature where lattice QCD backed by relativistic heavy ion experiments is providing valuable insight into hot medium, the situation is drastically different for cold hadronic matter at a density a few times that of the ordinary nuclear matter relevant for compact stars.  While asymptotic freedom should allow perturbative QCD to make well-controlled predictions at superhigh density, at the density regime relevant for compact stars, there are presently neither reliable theoretical tools nor experimental guides available to make clear-cut statements. The lattice method, so helpful in high-T matter, is hampered by the sign problem and cannot as yet handle the relevant  density regime.

What is generally accepted at the moment is that effective field theories formulated in terms of hadronic variables, guided by a wealth of experimental data, can accurately describe baryonic matter up to nuclear matter density $n_0\approx 0.16$ fm$^{-3}$ and perturbative QCD unambiguously predicts that color superconductivity should take place in the form of color flavor locking (CFL) at some asymptotically high density $n_{CFL}$~\cite{cfl-rev}. In between, say, $n_0\leq n \leq n_{CFL}$, presently available in the literature  are a large variety of model calculations which however have not been checked by first-principle theories or by experiments. The model calculations so far performed paint a complex landscape of phases from  $n_0$ to  $n_{CFL}$, starting with kaon condensation at $n_c^K \sim  3n_0$~\cite{BLR07}, followed by a plethora of color superconducting quark matter with or without color flavor locking near and above  the chiral restoration $n_c^{\chi SR}$ and ultimately CFL with or without kaon condensation. It is unclear which of the multitude of the phases could be realized and how they would manifest themselves in nature.

In this note, I would like to zero in near the chiral restoration
point $n_c^{\chi SR}$ and uncover a hitherto unsuspected novel
phenomenon that could take place very near $n_c^{\chi SR}$ in the
chiral limit. Treating dense nuclear matter in terms of skyrmion
matter, I will argue that at $n_p < n_c^{\chi SR}$, a skyrmion in
dense matter fractionizes  into two half skyrmions with chiral
$SU(N_f)\times SU(N_f)$ symmetry restored but with a non-vanishing
pion decay constant, and that in the presence of vector mesons, the
symmetry ``swells" to $SU(N_f)^4$ as the gauge (vector meson)
coupling $g\rightarrow 0$ near $n_c^{\chi SR}$ as predicted by
Georgi~\cite{georgi} and Harada and Yamawaki~\cite{HY:PR}. I will
conjecture that the skyrmion-half-skyrmion transition at $n_p$ is an
analog to what is referred to as ``deconfined quantum critical
phenomenon" in condensed matter physics~\cite{senthiletal} and
identify the phase $n_p\lsim n\lsim n_c^{\chi SR}$ with the
``hadronic freedom" regime and $n_p$ as the ``flash density," both
of which play an important role in describing  dense matter near and
just below the chiral transition point. I should stress that should
such a phase really exist, it could mean a significant deviation
from normal Fermi liquid and could therefore influence the possible
formation of color superconductivity as in high T superconductivity.

\sect{Half skyrmions and pseudogap phase:} As a way of approaching dense hadronic matter, I will adopt a skyrmion construction. This is chiefly because  a skyrmion with winding number $B$ is to encode large $N_c$ QCD for systems with $B$ baryons. This means that the skyrmion has the potential to provide a {\it unified approach} to baryonic dynamics, not only that of elementary baryon but also the structure of complex nuclei as well as infinite matter, the power and versatility that are missing in other approaches. Even more intriguingly, the CFL phase can also be described as a skyrmion matter of different form --  referred to as ``superqualiton" matter. Thus the transition from normal matter to CFL matter can be considered as a skyrmion-superqualiton  transition mediated by half-skyrmions in between.

Up to date, most of the works done on skyrmions relied on the Skyrme Lagrangian that contains the  current algebra term and the Skyrme term, viz,
\be
{\cal L}= \frac{F_\pi^2}{4} {\Tr} (\del_\mu U\del^\mu U^\dagger)+\frac{1}{32e^2}{\Tr} [U^\dagger\del_\mu U, U^\dagger\del^\nu U]^2
 \ee
 implemented with mass terms. But there are reasons to believe that vector degrees of freedom are essential for reliably describing systems with $B > 1$~\cite{MR-Yukawa}. Indeed, the recent development in holographic dual QCD (hQCD) clearly shows that the infinite tower of vector mesons encapsulated in five-dimensional (5D) Yang-Mills Lagrangian drastically modify the structure of the baryon arising as an instanton~\cite{HRYY}. In fact, not only generic soliton properties such as its topological stability but also chiral dynamics of nucleons including their electroweak form factors are strongly affected by the tower of vector mesons. This indicates that dense matter should be described with a hidden local symmetric Lagrangian with the infinite tower. However at present, the holographic dual Lagrangian approach is valid in the large $N_c$ and large 't Hooft ($\lambda =g_{YM}^2 N_c$) limit and making subleadiing corrections -- which are needed for treating dense medium -- appears to be extremely difficult. In incorporating vector degrees of freedom, I will therefore adopt Harada-Yamawaki HLS Lagrangian which can be viewed  as a truncation of the infinite tower with the higher lying vector mesons integrated out~\cite{HMY}. The effects of the integrated-out vectors will then be lodged in the parameters of the truncated Lagrangian that involves only the ground-state vectors, coupled chiral invariantly to the pions.

I will start with the skyrmion matter constructed with the Skyrme Lagrangian and then go over later to a hidden local symmetric Lagrangian containing the lowest lying vector mesons $\rho$ (and $\omega$). There have been a series of works on dense matter treated with the Skyrme Lagrangian~\cite{Skyrme-matter} on which I will base my beginning arguments.

In \cite{Skyrme-matter}, following the seminal work of Klebanov~\cite{klebanov}, density effect is simulated by putting skyrmions on crystal and squeezing the crystal. In (3+1) dimensions, it is found to be energetically favorable to arrange the skyrmions as a face-centered cubic crystal (FCC) lattice~\cite{FCC}. One should however recognize that there is no proof that this is indeed the absolute minimal configuration. There may be other configurations that are more favorable. Indeed, it has been recently shown that in baby-skyrmion systems~\cite{karliner}, of all possible crystalline structures, it is  the hexagonal, not the cubic, that  gives the minimal energy.   This caveat notwithstanding,  I will base my discussions on the FCC crystalline structure.   I will say more on this below, in particular concerning certain qualitative features that could be different for different crystalline structures.

Briefly, what is done in \cite{Skyrme-matter} is as follows. The crystal configuration made up of skyrmions has each FCC lattice site occupied by a single skyrmion centered with $U_0=-1$ with each nearest neighbor pair relatively rotated in isospin space  by $\pi$ with respect to the line joining the pair. In order to have the Skyrme Lagrangian possess the correct scaling under scale change of the crystalline, the dilaton scalar $\chi$ associated with the trace anomaly of QCD has to be implemented as suggested in \cite{BR91}. The energy density of the lattice skyrmions is then given by~\cite{Skyrme-matter}
\be
\epsilon &=& \frac{1}{4} \int_{Box} d^3x \left\{ \frac{f^2_\pi}{4} \left(
\frac{\chi_0}{f_\chi} \right)^2 \mbox{Tr}(\partial_i U_0^\dagger
\partial_i U_0) + \cdots
\right.
\nonumber \\
&& \left. \hskip 1em
+ \frac12 \partial_i \chi_0 \partial_i \chi_0 + V(\chi_0) \right\}
\label{EoverB}\end{eqnarray}
where $f_\pi$ is the physical pion decay constant (which is equal to the parametric constant $F_\pi$ at the tree order).
Here, the ellipsis stands for the familiar Skyrme quartic term and quark mass terms which need not be explicited, the subscript `box' denotes that the integration is over a single FCC box and the factor 1/4 in front appears because the box contains baryon number four. The subscript ``0" denotes the mean field, $f_\chi$ is the $\chi$ decay constant and $V(\chi)$ is the dilaton potential that encodes the ``soft" component of the trace anomaly associated with the spontaneous breaking of scale invariance~\cite{BHLR06}. The field $\chi$ is coupled to the chiral field $U$, so the mean field $ \chi^*=\la \chi\ra_n$ (where the asterisk indicates in-medium quantity with $n\neq 0$)  scales with the background provided by the crystal configuration. This behavior  mimics the ``intrinsic density dependence (IDD)"  required by Wilsonian matching of HLS theory to QCD at some matching scale~\cite{HY:PR}. The minimization of this energy density with respect to the coefficients of the Fourier expansion of the (mean) fields taken as variational parameters reveals that at some minimum size of the box corresponding to a density, say, $n_p$  of the matter, there is a phase transition from the FCC crystal configuration of skyrmions into a body-centered cubic crystal (BCC) configuration of half skyrmions as predicted on symmetry grounds~\cite{goldhaber-manton,FCC}. I should point out  two aspects here that characterize the transition. One is that what is involved here is a topology change, also observed  in (2+1) dimensions. Therefore it has the possibility of being stable against quantum fluctuations. The other is that in terms of the mean chiral field $U_0(x)=\sigma (x) +i\tau\cdot\pi$, the expectation value $\la\sigma\ra \propto \la\bar{q}q\ra$ is zero at $n_p$, so the transition is a chiral restoration phase transition.

\begin{figure}[htb]
\centerline{\epsfig{file=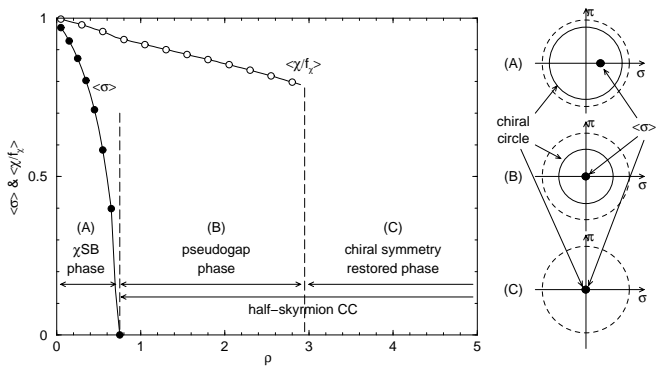,width=5cm,
angle=270}}
\caption{Average values of $\sigma=\frac12\mbox{Tr}(U)$ and
$\chi/f_\chi$ of the lowest energy crystal configuration at a given
baryon number density. Note that here $\rho$  stands for density
$n$.}\label{phase}
\end{figure}

The phase structure obtained numerically in \cite{Skyrme-matter} is shown in Fig.~\ref{phase} which summarizes what happens in the Skyrme model on crystal.
What is striking in this result is that while chiral symmetry is restored at $n_p$, the pion decay constant given by $f_\pi^*/f_\pi\propto \la\chi\ra_{n\geq n_p}/f_\chi\neq 0$. In terms of the chiral order parameter $\la\bar{q}q\ra\sim \la c{\Tr} (U+U^\dagger)\ra$, the half-skyrmion phase has $\la{\Tr}(U+U^\dagger)\ra=0$ but $c\sim f_\pi^* \neq 0$. A similar property has been proposed for high temperature and identified with a ``pseudogap phase" in analogy to high T superconductivity~\cite{zarembo}.  Although the connection is not clear, the phase --  phase (B) in Fig.~\ref{phase} -- between $n_p$ and the density denoted $n_c^{\chi SR}$ at which $f_\pi^*=0$ is called ``pseudogap phase." The density range of the pseudogap phase depends on the mass of the scalar $\chi$. It is possible that the range can be shrunk to a point for certain value of the mass but I will assume that there is such a finite window in which the pseudogap structure persists.

\sect{With hidden local symmetry:}
So far I have been discussing dense half-skyrmion matter built without vector meson degrees of freedom. The structure of the half-skyrmion phase becomes much more interesting and richer when vector mesons are present. As mentioned above,  the skyrmion structure with an infinite tower of vector mesons is totally different from the skyrmion with pion field only~\cite{HRYY}: even with only the ground state vector meson gauged into the Skyrme model, the basic feature of the resulting skyrmion gets drastically modified~\cite{brihayeetal}. Furthermore such a modification seems to be required if the skyrmion picture is to represent nature quantitatively~\cite{MR-Yukawa}.

There are basically two ways to introduce vector mesons into the skyrmion matter. They are essentially equivalent provided that local gauge invariance is implemented. I will resort to what I would call ``bottom-up" approach, that is to generate vector bosons as emergent gauge fields~\cite{HY:PR}. I will comment on the other, ``top-down" approach later.

The idea is simply that the chiral field $U=e^{2i\pi/F_\pi}$ can be written in terms of the left and right coset-space coordinates as
 \be
U=\xi_L^\dagger\xi_R\label{Up}
 \ee
with the transformation under $SU(N_f)_L\times SU(N_f)_R$ as
$\xi_L\rightarrow \xi_L L^\dagger$ and $\xi_R\rightarrow \xi_R R^\dagger$ with
$L(R)\in SU(N_f)_{L(R)}$.  Now the redundancy that is hidden, namely, the invariance under the $local$
transformation
 \be
\xi_{L,R}\rightarrow h(x)\xi_{L,R}
 \ee
where $h(x)\in SU(N_f)_{V=L+R}$ can be elevated to a local gauge invariance~\cite{HY:PR} with the corresponding gauge field
$V_\mu\in SU(N_f)_{V}$ that transforms
 \be
V_\mu\rightarrow h(x)(V_\mu-i\del_\mu)h^\dagger (x).
 \ee
(If one parameterizes $\xi_{L,R}=e^{i\sigma/F_\sigma}e^{\mp i\pi/F_\pi}$, gauge-fixing with $\sigma=0$ corresponds to unitary gauge, giving  the usual gauged nonlinear sigma model with a mass term for the gauge field.) The resulting  HLS Lagrangian takes the form \cite{georgi} (with $V_\mu=g\rho_\mu$):
 \be
{\calL}&=& \frac{F_\pi^2}{4}{\Tr}\{|D_\mu\xi_L|^2 + |D_\mu\xi_R|^2\nonumber\\
&& + \kappa |D_\mu U|^2\} \ - \frac{1}{2} \,
\mbox{Tr} \left[ \rho_{\mu\nu} \rho^{\mu\nu} \right] +\cdots \label{hls1}
 \ee
 where the ellipsis stands for higher derivative and other higher dimension terms including the gauged Skyrme term. ( I should mention parenthetically that the above construction can be extended to an infinite tower of vector mesons spread in energy in the fifth dimension, leading to a ``dimensionally deconstructed QCD" which is encapsulated in a 5D Yang-Mills theory~\cite{son-stephanov}. The latter is essentially equivalent in form to  the 5D Yang-Mills theory of holographic dual QCD that comes from string theory~\cite{HRYY}.)

 No satisfactory  construction of dense skyrmion matter with the Lagrangian (\ref{hls1}) exists at present~\cite{PRV-footnote}. However it is not difficult to see what happens in the presence of HLS degrees of freedom. As in the case of the Skyrme crystal, we expect that the skyrmion will fractionize at some $n_p$ into two half skyrmions, one given by $\xi_L^\dagger$ and the other by $\xi_R$. We can think of the former as ``left-half-skyrmion" and the latter ``right-anti-half-skyrmion," which are bound to a single skyrmion.  This resembles
  ``deconfined quantum critical phenomenon" in  (2+1) condensed matter systems, e.g.,  the transition from  the N\'eel magnetic phase to the VBS paramagnetic phase characterized by the deconfinement of a baby-skyrmion into two half baby-skyrmions~\cite{senthiletal}. In this condensed matter example where an ``emergent" $U(1)$ gauge field plays a crucial role, half-skyrmions are confined (or bound) to skyrmions in the initial (N\'eel) and final (VBS) phases and the transition occurs via the fractionized (or deconfined) half-skyrmion phase.

 \sect{Transition from nuclear matter to CFL phase:}
 It is tempting to think of the phase transition from normal baryonic matter to quark matter going via the half-skyrmion phase as in the condensed matter case. To see whether this analogy can be made closer, let us consider the CFL phase of quark matter. In the real world of two (u and d) light flavors and one heavy (s) flavor,  a variety of model calculations predict a multitude of superconducting states, some unstable and some others (such as LOFF crystalline) presumably stable, but I am going to consider, for simplicity, the CFL configuration which is favored for degenerate quark masses. Furthermore, there is also a possibility that the CFL phase can come down in density all the way to the nuclear matter density for a light-enough s-quark mass, say, in the presence of strong $U(1)_A$ anomaly~\cite{CFL-anomaly}.

 Since in the CFL phase, the global color symmetry $SU(3)$ is completely broken and chiral $SU(3)_L\times SU(3)_R$ (for $N_f=3$) is broken down to the diagonal subgroup $SU(3)_V$~\cite{cfl-rev}, low-energy excitations can be described by the coordinates $\xi_{C+L}\in SU(3)_{C+L}$ and $\xi_{C+R}\in SU(3)_{C+R}$ given in terms of the octet pseudoscalar $\pi$ and  the octet scalar $s$. The scalars are eaten up by the gluons which become massive and map one-to-one to the vector mesons present in the hadronic sector. The Lagrangian that describes low-energy excitations is of the same local gauge invariant form as (\ref{hls1}). The gauge symmetry here is explicit, not hidden as in the hadronic sector  but I will nonetheless call it HLS'. Now as in the hadronic sector, the HLS' Lagrangian supports solitons that carry fermion number $B$, which are nothing but skyrmions~\cite{qualiton}. The CFL soliton is called ``superqualiton" to be distinguished from the soliton in the hadronic phase.  It is actually a quark excitation on top of the vacuum with condensed Cooper pairs, effectively color singlet with spin 1/2. But in this formulation, it is a topological object.

 Given the skyrmion matter for $n\lsim n_p$ and the superqualiton matter for $n\gsim n_c^{\chi SR}$, the transition from nuclear matter to CFL matter can be considered as a skyrmion-superqualiton transition with half skyrmions figuring in between. This is the analogy to the N\'eel-VBS transition with half-skyrmions (spinons) at the boundary.
 Independently of whether this analogy is just a coincidence or has a non-trivial meaning, what is significant is that the pseudogap region can deviate strongly from the Fermi-liquid state that is usually assumed in studying color superconductivity.

\sect{Vector symmetry and hadronic freedom at high density:}
 What is perhaps the most significant for dense matter near chiral restoration  is that the half-skyrmion (or pseudogap) state exhibits an emerging or ``enhanced" symmetry. In HLS theory, the half-skyrmion state has the chiral $SU(N_f)_L\times SU(N_f)_R$  restored. Thus  for $n_p\leq n < n_c^{\chi SR}$,
 \be
 (F_\sigma/F_\pi)^2\equiv a=1, \ \  F_\pi\neq 0,
 \ee
 which corresponds to $\kappa=0$ in Eq.~(\ref{hls1}). Note however that the gauge coupling $g\neq 0$, so the vector meson remains  massive. Since the vector meson  is massive, $F_\sigma$ is the decay constant for the longitudinal component of the vector meson, not of a free scalar. The gauge coupling $g$ goes to zero, however, at chiral restoration, $n=n_c^{\chi SR}$. This corresponds to Georgi's ``vector limit."  As noted by Georgi~\cite{georgi}, at this point the symmetry ``swells" to $SU(N_f)^4$, with $\xi_L$ and $\xi_R$ transforming under independent $SU(N_f)\times SU(N_f)$ symmetries~\cite{wardidentity},
 \be
 \xi_L\rightarrow h_L(x) \xi_L L^\dagger, \ \  \xi_R\rightarrow h_R(x) \xi_R R^\dagger,
 \ee
 where ${L,R}$ and $h_{L,R}$ are the unitary matrices generating the corresponding global and local $SU(N_f)$ groups. The hidden local symmetry is the diagonal sum of $SU(N_f)_{h_L}$ and $SU(N_f)_{h_R}$. Away from the vector limit, the non-zero gauge couplings break the vector symmetry explicitly producing the nonzero vector meson mass and couplings for the transverse components of the vector mesons. In terms of this symmetry pattern, we see that the pseudogap phase is the regime where one has $a=1$ ($\kappa=0$) and weak gauge coupling $g\rightarrow 0$. I note that this is precisely the regime in which ``hadronic freedom"   is operative. This allows us to identify the onset density for the pseudogap phase $n_p$ with the ``flash density" $n_{flash}$  -- the density counterpart of the ``flash temperature" $T_{flash}$.

The pseudogap region between $n_{flash}$ and $n_c^{\chi SR}$ has an
important astrophysical implication. With $\kappa\rightarrow 0$
($a\rightarrow 1$), the gauge coupling $g$ goes to zero
as density approaches $n_c^{\chi SR}$, so hadrons interact weakly in
that regime. As in matter at high temperature between
$T_{flash}\simeq 125$ MeV and $T_c^{\chi SR}\simeq 175$ MeV, it can
be viewed as a region of hadronic freedom.  In the case of high
temperature, say, in peripheral collisions of heavy-ion collisions,
nearly non-interacting light mesons making up $\sim 32$ degrees of
freedom flow freely from $T_c^{\chi SR}$ down to the flash
temperature $T_{flash}$  at which they go on shell and become
strongly interacting. The observed $\rho^0/\pi^-$ ratio in the STAR
Au-Au collisions -- which is difficult to understand in standard
approaches -- can be simply explained with this
mechanism~\cite{BLR-STAR}.  In \cite{BLPR-kcond}, a similar
reasoning was made to predict kaon condensation at a density $\sim 3
n_0$. There the assumption was that kaons must condense somewhere
between the flash density $n_{flash}$ and $n_c^{\chi SR}$, most
likely closer to the latter. Therefore one can start from the vector
manifestation fixed point of HLS theory with $a= 1$ and $g=0$ but
$F_\pi\neq 0$. This calculation reinforced the previous conclusion
that kaons must condense {\em before} any other phase changes can
take place and hence determine the fate of compact stars. This is
reviewed in \cite{BLR07}.

\sect{Further remarks:} The main assumption made in this note is that dense
baryonic matter can be simulated on crystal using HLS
Lagrangian. There are several questions one can raise here.

The first is whether there are no other crystal configurations that could (1) give a lower ground state and (2) induce different skyrmion fractionization. The answer to this is not known. It is unquestionably an important question to address. For instance,  in (2+1) dimensions, while for the known square-cell configuration, a baby-skyrmion fractionizes into two half-skyrmions, it is the hexagonal configuration that has the minimal energy and induces the fractionization of a baby-skyrmion into four {\em quarter-skyrmions}~\cite{karliner}.

Given that the skyrmion-half-skyrmion transition scenario is anchored on the crystalline structure at the mean field level, one wonders whether quantum fluctuations would not wash out the soliton structure of the half-skyrmion matter. As mentioned in \cite{Skyrme-matter}, since nuclear matter is known to be a liquid, not a crystal, it might be that quantum fluctuations would ``melt" the crystal. The phase change could then  be merely a lattice  artifact. In addition,  the spin and statistics of the half-skyrmion would require quantization. It seems highly plausible however that given that the transition involved here is a topology change, the phase change be robust against quantum fluctuations. Similar issues are raised in condensed matter physics where the concept of ``topological order" is invoked for robustness of topology-changing phase transitions.

The next unanswered question is the mechanism for the
fractionization of a skyrmion to two half-skyrmions at $n_p$. The
fractionization at some density seems generic, taking place both in
(2+1) and (3+1) dimensions. The treatment made in this note was
based on energetics considerations but the mechanism was left
unclarified. In the condensed matter case discussed in
\cite{senthiletal}, the key role for the fractionization is played
by the emergent $U(1)$ gauge field and its monopole structure.  The
pair of half-skyrmions (referred to as ``up-meron" and
``down-anti-meron" in \cite{senthiletal}) are confined -- or bound -- to a single skyrmion
in both the initial N\'eel state and the final VBS state but the
skyrmion fractionizes into half skyrmions at the boundary due to the
``irrelevance" of the monopole tunneling, with an emergent global symmetry not present in the many-body Hamiltonian. It would be exciting to
see  a similar mechanism at work in the present case. It could
elucidate what the hadronic phase could be at the doorway to color
superconductivity, should the latter survive the Brown-Bethe scenario for black-hole collapse following
kaon condensation~\cite{BLR07}. In this regard, it would be
interesting to investigate the skyrmion-half-skyrmion transition in
terms of the instantons and merons of  5D Yang-Mills Lagrangian of
hQCD which would reveal the role of the infinite tower.

If the pseudogap phase is indeed the hadronic freedom region, how
can one exploit the background provided by the half-skyrmion matter
for describing kaon condensation? One may embed and bind $K^{-}$'s in  dense
half-skyrmion matter where $a\approx 1$ and $g\sim 0$ and exploit
that  in compact stars, electrons with high chemical potential decay
into $K^-$'s once the kaon mass falls sufficiently low and the kaons
Bose-condense. To do this calculation, it may be necessary to know
what the quantum structure of the half-skyrmion phase is.


\vspace{-0.4cm}

\begin{thebibliography}{99}
\bibitem{cfl-rev} For review, see K. Rajagopal and F. Wilczek, {\it At
the frontier of particle physics: Handbook of QCD}\ ed by M. Shifman
(World Scientific, Singapore, 2001) Vol. 3, p.2061.

\bi{BLR07} For review, see G.E. Brown, C.-H. Lee and M. Rho, ``Recent developments on kaon condensation and its astrophysical implications,"
Phys. Rept., in press, e-Print: arXiv:0708.3137 [hep-ph].

\bi{georgi} H. Georgi, ``New realization of chiral symmetry,"  Phys.\ Rev.\ Lett. {\bf 63}, 1917 (1989);  ``Vector realization of chiral symmetry,"
Nucl.\ Phys. {\bf B331},  311 (1990).

\bibitem{HY:PR} M. Harada and K. Yamawaki, ``Hidden local symmetry at loop: A new perspective of composite
gauge bosons and chiral phase transition,"   Phys. Rept. {\bf 381}, 1 (2003).

\bibitem{senthiletal} T. Senthil et al., ``Deconfined quantum critical points," Science {\bf 303}, 1490 (2004).

\bi{MR-Yukawa} M. Rho,  ``Hidden local symmetry and the vector
manifestation of chiral symmetry in hot and/or dense matter,"
{Prog. Theor. Phys. Suppl.}  {\bf 168},  (2007) 519. A glaring
defect of the skyrmion with chiral fields only and with no other
fields (such as the vector mesons $\rho$, $\omega$ etc) is that when
applied to nuclei, the parameters needed to even approximately fit
nature are totally unnatural. For instance, the parameter $f_\pi$ is
much too small compared with the physical value $f_\pi\approx 93$
MeV -- this is so even for a single nucleon -- and the pion mass
parameter $m_\pi$ is much too large compared with its free-space
value $m_\pi\approx 140$ MeV. See e.g. R. Battye and P. Sutcliffe,
\pr\ {\bf C73}, 055205 (2006)  and R. Battye et al, hep-th/0605284.
When the parameters are taken to be close to their physical values,
the resulting structure at the mean field level of complex nuclei,
e.g., shape, comes out to be completely different from what is known
in nature.

\bi{HRYY} D.-K. Hong, M. Rho, H.-U. Yee and P. Yi,  ``Chiral dynamics of baryons from string theory," Phys. Rev.  {\bf D76},  061901 (2007);  ``Dynamics of baryons from string theory and vector dominance,"  { JHEP} {\bf 09}, 063 (2007); ``Nucleon form factors and hidden symmetry in holographic QCD," arXiv:0710.4615 [hep-ph].

\bi{HMY} M. Harada, S. Matsuzaki and K. Yamawaki, `` Implications of holographic QCD in ChPT with hidden local symmetry,"  {\pr}\ {\bf D74}, 076004 (2006).

\bi{Skyrme-matter}  B.-Y. Park, D.-P. Min, M. Rho and V. Vento, \np\ {\bf A707}, 381 (2002);  H.-J. Lee, B.-Y. Park, D.-P. Min, M. Rho and V. Vento, \np\ {\bf A723}, 427 (2003).

\bi{klebanov} I.R. Klebanov,  ``Nuclear matter In the Skyrme model, " Nucl. Phys. {\bf B262}, 133.

\bi{FCC} M. Kugler and S. Shtrikman,  ``Skyrmion crystals and their symmetries,"   {Phys. Rev.}\ {\bf D40}.  3421 (1989);  L. Castellojo, P.S.J.  Jones, A.D. Jackson and J.J.M  Verbaarschot,  ``Dense skyrmion systems," Nucl. Phys. {\bf A501}, 801 (1989).

\bi{karliner} I. Hen and M. Karliner, ``Hexagonal structure of baby skyrmion lattice," arXiv:0711.2387 [hep-th].

\bi{BR91} G.E.  Brown and M. Rho, ``Scaling effective Lagrangians in a dense medium,"  {Phys. Rev. Lett.} {\bf 66},  2720 (1991).

\bibitem{BHLR06} G.E. Brown, J.W.  Holt, C-H. Lee and  M. Rho, ``Late hadronization and matter formed
at RHIC: Vector manifestation, BR scaling and hadronic freedom," { Phys. Rept.} {\bf 439},  161 (2006).

\bi{goldhaber-manton} A.S. Goldhaber and N.S. Manton, `` Maximal symmetry of the Skyrme crystal,"  { \pl}  {\bf B198},  231 (1987).

\bi{zarembo} K. Zarembo, ``Possible pseudogap phase in QCD,"  JETP Lett.{\bf 75}, 59 (2002).

\bi{brihayeetal} Y. Brihaye, C.T. Hill and Zachos, ``Bounding gauged skyrmion masses," {\pr} {\bf D70}, 111502 (2004).

\bibitem{son-stephanov} D.T.  Son and M.A. Stephanov, `` QCD and dimensional deconstruction,"  {\pr}\
{\bf D69}, 065020 (2004).

\bi{PRV-footnote} An early effort was made in \cite{PRV-vector} to incorporate  the vector mesons $\rho$ and $\omega$ into skyrmion matter extending the construction of \cite{Skyrme-matter}. However the anomalous sector of the simplified Lagrangian used there was not consistent with the full vector dominance implied by both holographic dual QCD and hidden local symmetry structure of Harada-Yamawaki theory.  Therefore the results obtained there cannot be trusted as they stand. Work is in progress to correctly implement vector mesons in skyrmion physics~\cite{OMPR07}.

\bi{PRV-vector} B.-Y. Park, M. Rho and V. Vento, ``Vector mesons and dense skyrmion matter,"  Nucl. Phys. {\bf A736}:129 (2004).

\bi{OMPR07}  Y. Oh, D.-P. Min, B.-Y. Park and M. Rho, work in progress.

\bi{CFL-anomaly} T. Hatsuda,  M.  Tachibana, N. Yamamoto and G. Baym, ``New critical point induced by the axial anomaly
in dense QCD, " {\prl}\ {\bf 97},  122001 (2006).

\bi{qualiton}  D.K  Hong, M.  Rho and I.  Zahed,  ``Qualitons at high density," {Phys. Lett.} {\bf B468},   261 (1999).

\bi{wardidentity} In his formulation, Georgi considered that vector symmetry could encompass the situation where $g=0$ and $\kappa=0$ but $F_\pi\neq 0$ in what might be identified as a ``mended symmetry."  It is argued however in \cite{HY:PR} that with quantum corrections included, one cannot have $g=0$ and $f_\pi\neq 0$ (where $f_\pi$ is the physical pion decay constant) since this violates chiral Ward identity. A loop-hole to this no-go theorem could be that in medium, there are collective excitations and the enhanced symmetry could be an ``emergent" symmetry not visible in the fundmental theory, QCD.

\bi{BLR-STAR}   G.E. Brown, C-H. Lee and  M. Rho, ``Vector manifestation of hidden local symmetry, hadronic freedom, and the STAR $\rho^0 /\pi^-$  ratio,"
Phys. Rev. {\bf C74}, 024906 (2006).

\bi{BLPR-kcond} G.E. Brown, C-H. Lee, H.-J. Park and  M. Rho,  ``Study of strangeness condensation by expanding about the fixed point of the Harada-Yamawaki vector manifestation,"  Phys. Rev. Lett. {\bf 96}, 062303 (2006).




\end{thebibliography}
\end{document}